\documentclass[prd, 11pt,a4paper,groupedaddress,preprintnumbers,nofootinbib,floatfix, twocolumn, showpacs]{revtex4-1}

\usepackage[top=2.9cm, bottom=2.1cm, left=2cm, right=2cm]{geometry}
\usepackage[english]{babel}
\usepackage{amsmath,amssymb,amsfonts, dsfont}
\usepackage{graphicx}
\usepackage{color}
\usepackage{hyperref}
\usepackage{subfigure}
\usepackage{dcolumn}
\usepackage{bm}
\usepackage{float}
\usepackage{mathtools}
\usepackage{amsthm}
\usepackage{bbm}
\usepackage{pxfonts}
\usepackage[vcentermath]{youngtab}
\usepackage[all]{xy}
\usepackage{pstricks}
\usepackage{accents}
\usepackage{cases}
\usepackage{comment}
\usepackage{placeins}
\usepackage{xspace}
\usepackage{cancel} 
\usepackage{slashed}
\usepackage{soul}
\usepackage{feynmp}
\usepackage{multirow}
\usepackage{enumerate}

\newcommand{\betabeta}{\mbox{$(\beta \beta)_{0 \nu}  $}}
\newcommand{\be}{\begin{equation}}
\newcommand{\ee}{\end{equation}}

\newcommand{\Dmq}{\Delta m^2}
\newcommand{\eVq}{\ensuremath{\text{eV}^2}}

\begin{document}

\title{Neutrino masses and ordering via multimessenger astronomy}
\author{Kasper {\sc Lang\ae ble}}
\email{langaeble@cp3.sdu.dk}
\author{Aurora {\sc Meroni}}
\email{meroni@cp3.sdu.dk}

\affiliation{{CP}$^{ \bf 3}${-Origins} \& the Danish Institute for Advanced Study {\rm{Danish IAS}},  University of Southern Denmark, Campusvej 55, DK-5230 Odense M, Denmark.}
\author{Francesco {\sc Sannino}}
\email{sannino@cp3.dias.sdu.dk}
\affiliation{{CP}$^{ \bf 3}${-Origins} \& the Danish Institute for Advanced Study {\rm{Danish IAS}},  University of Southern Denmark, Campusvej 55, DK-5230 Odense M, Denmark.}

\begin{abstract} 
We define the theoretical framework and deduce the conditions under which multi-messenger astronomy can provide useful information about  neutrino masses and their ordering. The framework uses  time differences between the arrival of neutrinos and the other 
 light messenger, i.e. the graviton,  emitted in astrophysical catastrophes.     
We also provide a preliminary feasibility study elucidating the experimental reach and challenges for planned neutrino detectors such as Hyper-Kamiokande as well as future several megaton detectors. 
This study shows that future experiments can be useful in testing independently the cosmological bounds on absolute neutrino masses.
 Concretely the success of such measurements depends crucially on the available rate of astrophysical events and further requires development of high resolution timing  besides the advocated need of megaton size detectors.
\vskip .1cm
{\footnotesize  \it Preprint: CP$^3$-Origins-2016-010 DNRF90 \& DIAS-2016-10}
\end{abstract}

\pacs{04.80.Nn, 14.60.Lm, 14.60.St}
\maketitle
\newpage

\section{Introduction}

The fascinating discovery by the LIGO collaboration \cite{Abbott:2016blz} of
 ripples in the fabric of space-time ---the gravitational waves, first anticipated by A. Einstein a century ago--- 
shows us a completely new way of exploring the Universe.  
Gravitational waves (GW)s carry detailed information about astrophysical catastrophes and can provide a clear reference time for multi-messenger astronomy.
 In the next decade we therefore expect great advances from the experimental particle physics searches, on Earth and in space. 

It is therefore timely to ask whether it is possible to use these new extraordinary experimental achievements as new tools to help settling some of the open issues in particle physics. 

We know that the Standard Model (SM)
cannot be the ultimate theory of Nature since the neutrino sector and dark matter  are not yet properly accounted for.  
In fact, the nature of the three light active neutrinos $\nu_i$ ($i=1,2,3$) with definite mass $m_i$ is unknown.  
To date, neutrinos can still be Dirac fermions if particle interactions conserve
some additive lepton number, e.g.  the total
lepton charge $L = L_e + L_{\mu} + L_{\tau}$. However, if the total lepton charge is violated, they can have a Majorana nature \cite{Bilenky:2001rz, Bilenky:1987ty}. 
The only feasible experiment, so far,  that can unveil the nature of massive neutrinos is neutrinoless double beta, \betabeta ~decay (see e.g. \cite{Rodejohann:2011mu} for a review).

Another pressing question to answer is how light are neutrinos
\footnote{For a discussion on  neutrino speed measurements on Earth see e.g. \cite{Krauss:1987me, Longo:1987gc, Adamson:2015ayc}.}.
Experimental evidence of neutrino oscillations, and thus the existence of at
least three neutrino states, force us to include them in the
SM and give them small mass differences \cite{Agashe:2014kda}. However oscillation experiments are not sensitive to their  masses. That their masses are tiny, when compared to other SM particles, comes from cosmology where an upper bound on the sum of the active neutrinos {$ \sum_i m_i <0.23$}~eV can be established  \cite{Ade:2013zuv}. 
More recently, more stringent limits have been obtained through the Lyman alpha forest power spectrum,  {$ \sum_i m_i <0.12$}~eV
\cite{Palanque-Delabrouille:2015pga}. These constraints will be further tested independently by other   experiments such as beta decay  and neutrinoless double beta decay experiments. 
Future large scale structure surveys like the approved EUCLID \cite{Laureijs:2011gra}, 
will allow to constrain  $\sum_i m_i$ down to $0.01$~eV  when combined with Planck data.

The enormous disparity between the neutrino masses and the ones 
of the charged leptons and quarks suggests that  
the neutrino masses might be related to the existence 
of a new fundamental mass scale in particle physics, associated 
with the existence of new physics beyond that predicted by the SM.
The so called {\it see-saw} mechanism \cite{seesaw} gives an appealing explanation of neutrino mass generation explaining at the same time the
smallness of their masses and of their possible Majorana nature, through the existence of heavier fermionic SM singlets. It can also serve as stepping stone for an explanation of the 
observed baryon asymmetry in the universe through leptogenesis \cite{LeptoG}.
\\
\begin{table*}[t]
  \begin{tabular}{|c|cc|cc|}
    \hline\hline
    & \multicolumn{2}{c|}{Normal Ordering }
    & \multicolumn{2}{c|}{Inverted Ordering }
    \\
    \hline
    & bfp $\pm 1\sigma$ & $3\sigma$ range
    & bfp $\pm 1\sigma$ & $3\sigma$ range
    \\
    \hline
    \rule{0pt}{4mm}\ignorespaces
    $\sin^2\theta_{12}$
    & $0.304_{-0.012}^{+0.013}$ & $0.270 \to 0.344$
    & $0.304_{-0.012}^{+0.013}$ & $0.270 \to 0.344$
    \\[3mm]
    $\sin^2\theta_{23}$
    & $0.452_{-0.028}^{+0.052}$ & $0.382 \to 0.643$
    & $0.579_{-0.037}^{+0.025}$ & $0.389 \to 0.644$
    \\[3mm]
    $\sin^2\theta_{13}$
    & $0.0218_{-0.0010}^{+0.0010}$ & $0.0186 \to 0.0250$
    & $0.0219_{-0.0010}^{+0.0011}$ & $0.0188 \to 0.0251$
    \\[3mm]
    $\Dmq_{21} [10^{-5}~\eVq]$
    & $7.50_{-0.17}^{+0.19}$ & $7.02 \to 8.09$
    & $7.50_{-0.17}^{+0.19}$ & $7.02 \to 8.09$
    \\[3mm]
    $\Dmq_{3\ell} [10^{-3}~\eVq]$
    & $+2.457_{-0.047}^{+0.047}$ & $+2.317 \to +2.607$
    & $-2.449_{-0.047}^{+0.048}$ & $-2.590 \to -2.307$
    \\[3mm]
    \hline\hline \end{tabular}
  \caption{  \label{tab:results}Three-flavor oscillation parameters from the fit to global
    data after the NOW~2014 conference performed by the NuFIT group~\cite{Gonzalez-Garcia:2014bfa}.
   The
    numbers in the 1st (2nd) column are obtained assuming NO (IO). Note that $\Dmq_{3\ell} \equiv \Dmq_{31} > 0$ for NO and
    $\Dmq_{3\ell} \equiv \Dmq_{32} < 0$ for IO.} 
\end{table*}
\\
The detection of GW150914  \cite{Abbott:2016blz} has already ignited the experimental neutrino community (see e.g. the null search results of ANTARES and ICE-CUBE \cite{Adrian-Martinez:2016xgn}), and the next-generation kilometer-scale laser-interferometric GW detectors such as aLIGO \cite{Harry:2010zz}, aVIRGO \cite{Accadia:2011zzc}, and KAGRA \cite{Somiya:2011np} will have strong impact on multi-messenger astronomy. 
\\
The goal of this work is to investigate whether experiments, making use of GW detection in combination with the associated neutrino (and photon) counterparts,  can make a dent in 
understanding the ordering of neutrino masses. 
\\
It has been established in the past literature 
\cite{Fargion:1981gg, Ryazhskaya:1992bs,Beacom:1998yb,Beacom:2000qy, Arnaud:2001gt, Raffelt:2002tu, Nardi:2003pr, Zuluaga:2005ah,Strumia:2006db}) and more recently in \cite{Nishizawa:2014zna},
that valuable information on the neutrino masses can be obtained  by investigating   the time delay between the observation of neutrinos and gravitational waves emitted in astrophysical events such as supernovae.
 In the meanwhile,  neutrino physics has entered the precision era with the determination of the  reactor mixing angle, $\theta_{13}$ \cite{An:2012eh, Ahn:2012nd}.
 Furthermore, different experiments, ranging from cosmological surveys to particle experiments, are constraining the absolute neutrino mass to be less than  0.1 eV.
 \footnote{In \cite{Arnaud:2001gt}  it was shown that using gravitational waves and neutrino burst events detected using SuperKamiokande or SNO, one could be sensitive to absolute neutrino masses in the range [0.75, 1.1]eV for distances of about 10 kpc.  These mass values are now outdated.}
  It is  therefore  timely to think about this subject in a new light.
\\
Despite the progress in neutrino physics, current experiments cannot yet decide on the neutrino mass ordering and their absolute mass scale. \\ 
In the following we  will  briefly review the current status of neutrino ordering and mixing. 
Next, we explore the conditions under which multi-messenger astronomy can reveal or constrain the neutrino mass ordering and absolute mass, concluding with a preliminary feasibility investigation. 

 \section{Neutrino orderings: current status}
 Current available neutrino oscillation data \cite{Gonzalez-Garcia:2014bfa} (see Table \ref{tab:results}) are compatible with  two types of neutrino mass spectra. These depend on the sign of $\Dmq_{3\ell}$ ($\ell=1,2$) and are summarised below:
 {\it  \begin{enumerate}[i)]
 \item Spectrum with normal ordering (NO):\vspace{-5pt}
		\begin{align*}
		 & m_1 < m_2 < m_3,\quad \Delta m^2_{31} >0,\quad \Delta m^2_{21} > 0,\\
		 & m_{2(3)} = (m_1^2 + \Delta m^2_{21(31)})^{1\over{2}\, };
		\end{align*}
 \item Spectrum with inverted ordering (IO):\vspace{-5pt}
\begin{align*}
 & m_3 < m_1 < m_2,\quad \Delta m^2_{32}< 0,\quad \Delta m^2_{21} > 0,\\
 & m_{2} = (m_3^2 + \Delta m^2_{23})^{1\over{2}},\\
 & m_{1} = (m_3^2 + \Delta m^2_{23} - \Delta m^2_{21})^{1\over{2}}\, .
 \end{align*}
 \end{enumerate}}
 \noindent It should be kept in mind that
 $\Delta m^2_{31}(NO) = |\Delta m^2_{32}(IO)|$,
 where the notation is self-explanatory.~%
 Depending on the value of the lightest neutrino mass,
 $m_{min}$, the neutrino mass spectrum can be:
 {\it
 \begin{enumerate}[i)]
 \item[a)] Normal Hierarchical (NH):\vspace{-5pt}
\begin{align*}
 & m_1 \ll m_2 < m_3,\phantom{\quad \Delta m^2_{31} >0,\quad \Delta m^2_{21} > 0,}\\
 & m_2 \cong (\Delta m^2_{21})^{1\over{2}}
 \cong 8.68 \times 10^{-3}~\mathrm{eV},\\
 & m_3 \cong (\Delta m^2_{31})^{1\over{2}}
 \cong 4.97\times 10^{-2}~{\rm eV};
\intertext{ \item[b)] Inverted Hierarchical (IH):\vspace{-5pt}}
& m_3 \ll m_1 < m_2,\\
& m_{1,2} \cong |\Delta m^2_{32}|^{1\over{2}}\cong 4.97\times 10^{-2}~\mathrm{eV};
\intertext{ \item[c)] Quasi-Degenerate (QD):\vspace{-5pt}}
& m_1 \cong m_2 \cong m_3 \cong m_0,\quad m_0  \gtrsim 0.1~\mathrm{eV},\\
& m_j^2 \gg |\Delta m^2_{31(32)}|,\quad j=1,2,3\, .
\end{align*}
\end{enumerate}
}
We denote solar and atmospheric square mass differences respectively,  $\Delta m^2_{21}$ and  $\Delta m^2_{3\ell}$.
The current cosmological bounds
are strongly disfavouring the degenerate regime. However these results should be further tested independently for example by beta-decay 
 and neutrinoless double beta decay experiments.

\section{Multi-messenger astronomy}
The detection of GWs is a crucial test of general relativity and, as already discussed in the literature, it is also important to deduce other relevant physical properties. This new information can be derived when comparing, for example, their propagation velocity with those of photons and neutrinos coming both from the same astrophysical source.  

\subsection{Set-up} Let's start by considering a potential observation of an astrophysical catastrophe.  Using  the same notation of \cite{Nishizawa:2014zna},  we denote  with $T_g\equiv L/v_g$,  $T_{\nu_i} \equiv L/v_{\nu_i}$ and $T_{\gamma} \equiv L/v_{\gamma}$  respectively the time of propagation of a GW, a given neutrino mass eigenstate and photons with group velocities $v_g,v_{\nu_i}$, and $v_{\gamma}$. Following Fig.~\ref{fig:times} 
a GW is emitted at the time  $t_{g}^E$ from a source at distance $L$ and detected on  Earth at $t_g$. Similarly, we have emission and detection times for photons and neutrinos. For instance, astrophysical catastrophes  like the merging of a  neutron star binary or the core bounce of a core-collapsed supernova (SN) are believed to follow this pattern. 
The  difference of the arrival times between the GWs and neutrinos, $\tau_{obs}\equiv t_\nu-t_g$,   or the GW and  a photon, $\tau_{obs}^\gamma\equiv t_\gamma -t_g$,  are both observables, which can be positive or negative for an early or late arrival of a GW. 
Typically the emission times of the three signals (GW, $\gamma$ and $\nu$) do not coincide\footnote{In alternative theories of gravity  the three particles under study --- photons, gravitons and neutrinos--- can couple to different  effective  metrics. In this case the Shapiro delay is not the same for the three signals \cite{Desai:2008vj}. In this work however we assume the same coupling to the metric for all the signals. }. 
For instance in the supernova explosion SN1987A \cite{SN},  the neutrinos arrived approximately 2 -- 3 hours before the associated photons.
\\
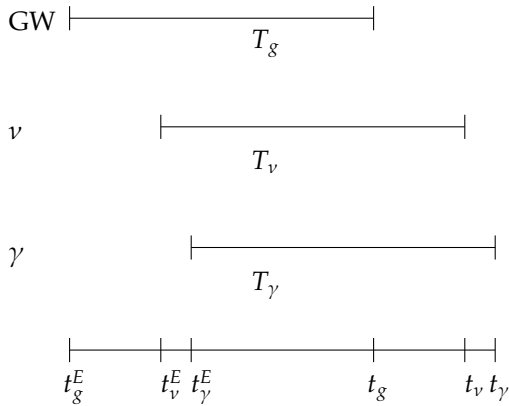
\begin{figure}[t]
	\unitlength=0.8mm
	\begin{picture}(120,80)
	\put(10,63){GW}
	\put(50,60){$T_{g}$}
	\put(20,65){\line(1,0){50}}
	\put(20,63){\line(0,1){4}}
	\put(70,63){\line(0,1){4}}
	\put(10,45){$\nu$}
	\put(50,40){$T_{\nu}$}
	\put(35,47){\line(1,0){50}}
	\put(35,45){\line(0,1){4}}
	\put(85,45){\line(0,1){4}}
	\put(10,25){$\gamma$}
	\put(50,20){$T_{\gamma}$}
	\put(40,27){\line(1,0){50}}
	\put(40,25){\line(0,1){4}}
	\put(90,25){\line(0,1){4}}
	\put(20,10){\line(1,0){70}}
	\put(20,8){\line(0,1){4}}
	\put(35,8){\line(0,1){4}}
	\put(40,8){\line(0,1){4}}
	\put(90,8){\line(0,1){4}}
	\put(70,8){\line(0,1){4}}
	\put(85,8){\line(0,1){4}}
	\put(19,3){$t_{g}^E$}
	\put(35,3){$t_{\nu}^E$}
	\put(40,3){$t_{\gamma}^E$}
	\put(69,3){$t_g$}
	\put(89,3){$t_\gamma$}
	\put(85,3){$t_\nu$}
	\end{picture}
	\caption{\label{fig:times}  GW, neutrino  and photon propagation in time.}
\end{figure}
\begin{figure*}[t]
	\subfigure
	{\includegraphics[width=7.5cm]{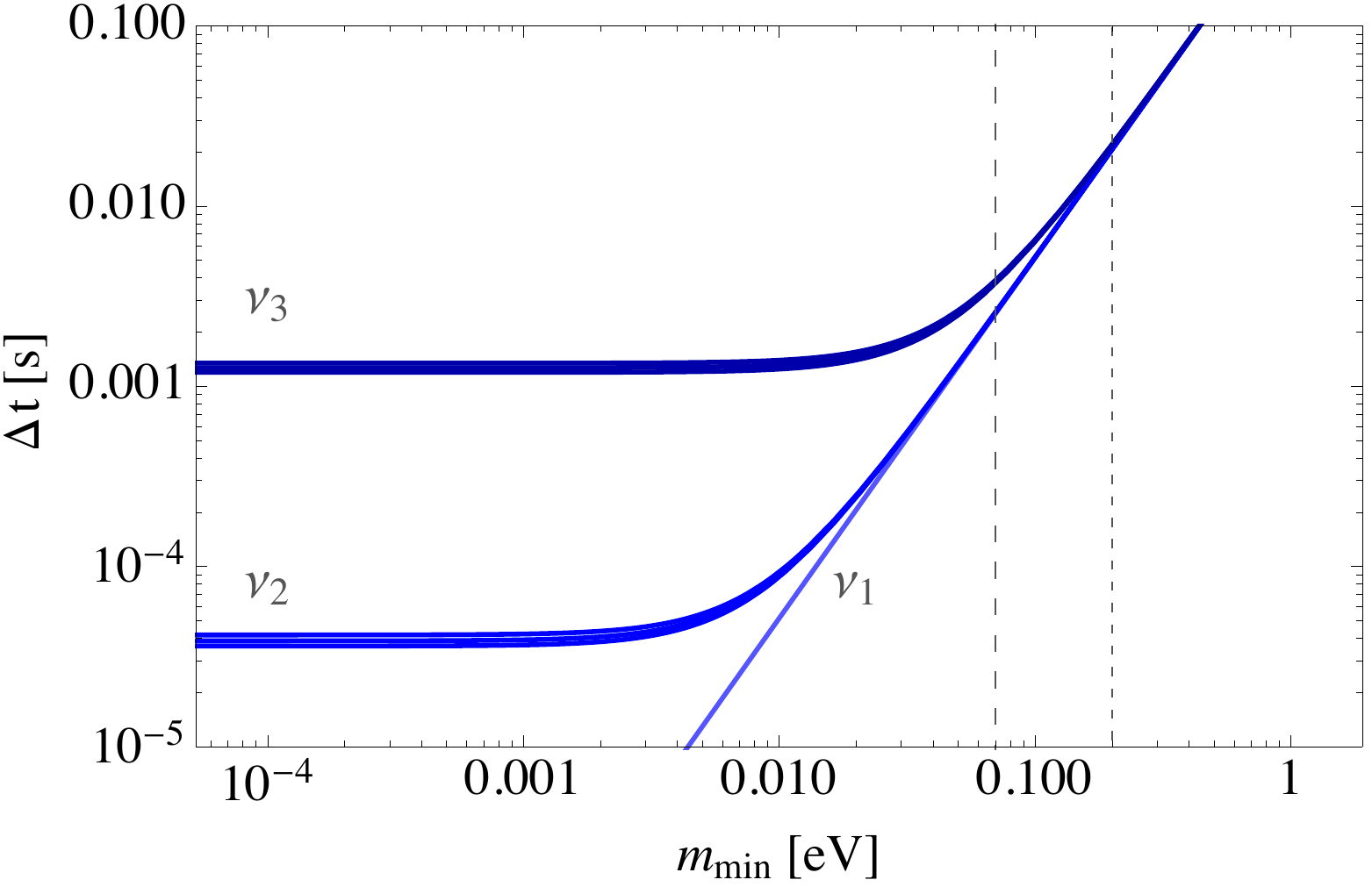}}
	\hspace{.6cm}\subfigure
	{\includegraphics[width=7.5cm]{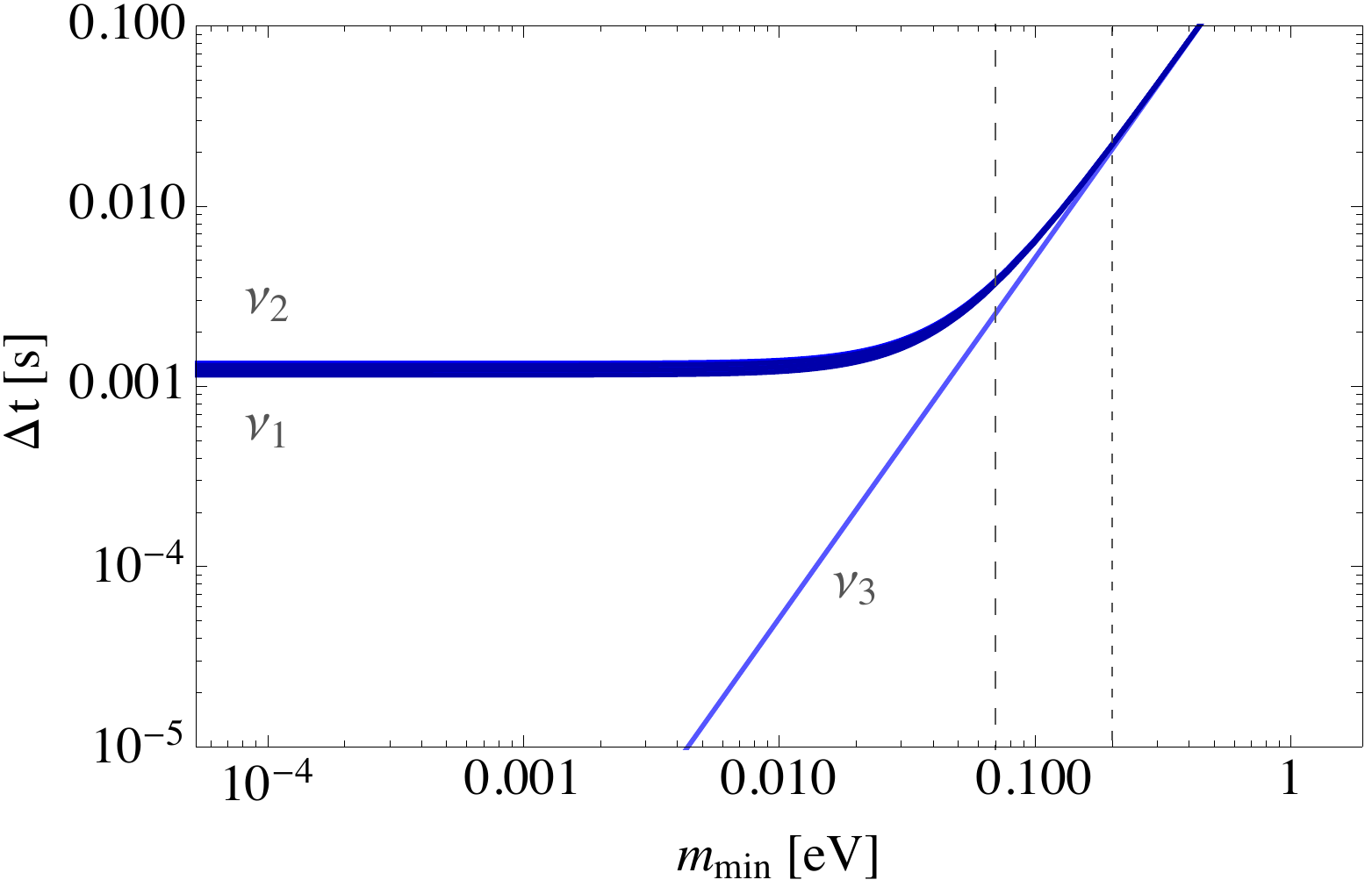}}
	\vspace*{-10pt}\caption{\label{fig:10MeV} The range of $\Delta t_i$ ($i=1,2,3$), the time delay of neutrinos with respect to photons, vs the lightest of the neutrino masses, $m_{min}$, for a distance of 1~Mpc and 10~MeV. 
		We show the results for NO and IO (left and right panels) considering a
		the 3$\sigma$ uncertainty in the oscillations parameters given in Table \ref{tab:results}. The dashed and  dotted  vertical lines correspond  to the Planck limit on the sum of neutrinos masses and the perspective upper limits from the KATRIN experiment (more details in the text). }
\end{figure*}
Let us assume now that a neutrino is emitted at $t_{\nu}^E=t_{g}^E + \tau^{\nu}_{int}$ and detected at time $t_\nu$.  A relativistic mass eigenstate neutrino with mass $m_i \, c^2\ll E$ ( $i=1,2,3$ ) propagates with a group velocity:
\be \frac{v_i}{c}=  1-\frac{m_i^2 c^4}{2 E^2}+\mathcal{O}\left(\frac{m_i^4 c^8}{8 E^4} \right)\, ,\ee
where we assumed that the different species of neutrinos have been produced with a common energy value $E$. 
If a given neutrino is produced by a source  at a distance $L$, the time-of-flight delay $\Delta t_i$ with respect to a massless particle, emitted by the same source at the same time,  is 
\be \Delta t_i 
\cong \frac{m_i^2 c^4}{2 E^2} \frac{L}{c} = 2.57 \left( \frac{m_i c^2}{\rm eV} \right)^2  \left( \frac{E}{\rm~MeV}  \right)^{-2} \frac{L}{50 {\rm kpc}}~s. \label{eq:TOF}\ee
Here we do not  take into account cosmic expansion since we consider sources at low redshift, $z \ll 0.1$.  This causes an error less than 5\%.
 From the expression in \eqref{eq:TOF} we observe that larger distances and small neutrino energies are needed in order to maximise the experimental sensitivity.
For distances around 50 kpc (SN1987A) and an energy of 10~MeV, a neutrino with a mass of 0.07 eV (the upper current absolute mass scale inferred from the Planck collaboration
 \cite{Ade:2013zuv}) would arrive   $\sim 10^{-4}$~s later than a massless particle. 
 Similar to \eqref{eq:TOF} we express the time delay between the arrival of two neutrino mass eigenstates as:
\be \Delta t_{\nu_i \nu_j}=\Delta t_i -\Delta t _j =\frac{\Delta m^2_{ij} c^4 }{2 E^2} T_0 \, ~~{\rm with} ~~ T_0=\frac{L}{c}\ , \label{eq:TOFH}\ee
with $\Delta m^2_{ij}= m_i^2-m_j^2$ and to leading order in $m^2 c^4/E^2$. {We note, that in this limit the time intervals do not depend on the absolute neutrino mass scale,}
{but  solely on the square mass differences which are determined experimentally (see Table \ref{tab:results})}.

\subsection{Disentangling neutrino mass ordering}
We are now equipped with the needed information to address our overarching quests. From  \eqref{eq:TOFH} we observe that if the detector uncertainty is  $10^{-3}$~s we are able to disentangle the  atmospheric (solar) squared mass differences with a signal coming from  a distance larger than 0.8~ (26)~Mpc   assuming neutrinos have an energy of about 10~MeV.   
 These distances decrease if we lower the neutrino energy. 
\\
This means, that for neutrinos with an average energy of 10~MeV,  the delay time of the heaviest neutrino mass eigenstate with respect to the lightest is larger than $10^{-3}$~s independently of the absolute neutrino mass scale and hierarchy, for distances larger than $\sim 0.8$ Mpc. Therefore, assuming an accuracy of $10^{-3}$~s, the relevant sources are those at distances larger than $0.8$~Mpc. With better time accuracy the distance decreases linearly. \\
We show in Fig.~\ref{fig:10MeV}  the time delay (for each mass eigenstate)  $\Delta t_i$  considering NO  and IO (left and right panels respectively) as function of the lightest neutrino mass, setting the neutrino energy to 10~MeV and the  distance of the source to 1~Mpc. 
The physically relevant arrival time differences between neutrino mass eigenstates $\Delta t_{\nu_i \nu_j}$ can be readily determined from Fig.~\ref{fig:10MeV}. 
 We also report in the plot the future sensitivity on the absolute neutrino mass of the $\beta$-decay experiment KATRIN~\cite{Robertson:2013ziv} which is expected to be around $0.2$~eV and  the  constraints given by the Planck Collaboration on the sum of the light active neutrinos \cite{Ade:2013zuv}
$
\sum_i m_i \leq 0.23~ \mbox{eV\, 95\% CL}.
$
In Table \ref{tab:BP} we produce relevant benchmark neutrino time lapses considering two different source-distances for different values of the lightest neutrino mass for 10~MeV neutrinos.  Table \ref{tab:BP5MeV}  shows the substantial gain in time-lapse for the  distances of 1  Mpc   but with a neutrino mass energy of 5~MeV which is still within experimental reach \cite{Abe:2011ts}. For illustrative purpose we give similar data for 10 Mpc in parenthesis.
\begin{table}[t]
	\begin{tabular}{|c|c|c|c|}
		\hline
		$m_{min}$ [eV] &   
		\multicolumn{2}{c|}{$\Delta t _{\nu_i}$ [s] }\\
		\hline \hline
		&  NO & IO\\
		\hline
		\multirow{ 3}{*}{0}      
		&  0 						       & $1.23 \cdot10^{-5} (10^{-3})$\\
		&   $3.86 \cdot10^{-7}(10^{-5})$         	 & $1.26 \cdot10^{-5} (10^{-3})$\\   
		&   $1.26 \cdot10^{-5}(10^{-3})$          	 & 0 \\ 
		\hline
		\multirow{ 3}{*}{0.01}   
		& $5.14 \cdot10^{-7}(10^{-5})$         	 & $1.28 \cdot10^{-5}(10^{-3})$\\
		&   $9.00 \cdot10^{-7}(10^{-5})$         	 & $1.32 \cdot10^{-5} (10^{-3})$\\   
		&   $1.32 \cdot10^{-5}(10^{-3})$          	 & $5.14 \cdot10^{-7}(10^{-5})$  \\ 
		\hline
	\end{tabular}
	\caption{  \label{tab:BP} Benchmark time lapses for $\nu_{1}$, $\nu_{2}$ and $\nu_{3}$ respectively. We consider a distance of 10 kpc (1 Mpc) and a neutrino energy of $E=10$~MeV.} 
\end{table}
\begin{table}[t]
	\begin{tabular}{|c|c|c|c|}
		\hline
		$m_{min}$ [eV] &   
		\multicolumn{2}{c|}{$\Delta t _{\nu_i}$ [s] }\\
		\hline \hline
		&  NO & IO\\
		\hline
		\multirow{ 3}{*}{0}    
		&  0 						       & $4.91 \cdot10^{-3} (10^{-2}) $\\
		&   $1.54 \cdot10^{-4}(10^{-3}) $         	 & $5.06 \cdot10^{-3} (10^{-2}) $\\   
		&   $5.06 \cdot10^{-3}(10^{-2})$          	 & 0 \\ 
		\hline
		\multirow{ 3}{*}{0.01}  
		& $2.06 \cdot10^{-4}(10^{-3}) $         	 & $5.11 \cdot10^{-3}(10^{-2}) $\\
		&   $3.60 \cdot10^{-4}(10^{-3}) $         	 & $5.27 \cdot10^{-3} (10^{-2})$\\   
		&   $5.27 \cdot10^{-3}(10^{-2})$          	 & $2.06 \cdot10^{-4}(10^{-3})$  \\ 
		\hline
	\end{tabular}
	\caption{  \label{tab:BP5MeV} Benchmark time lapses for $\nu_{1}$, $\nu_{2}$ and $\nu_{3}$ respectively. We consider a distance of 1 (10) Mpc and a neutrino energy of $E=5$~MeV.} 
\end{table}
\\
From Fig.~\ref{fig:10MeV}  we observe that for the given distance and energy, the NO and IO spectra differ by having different time delay patterns. We note that for IO the delay between the two heaviest mass eigenstates is equivalent to the time lapse between the first two lighter mass eigenstates for NO.
If we consider a conservative time accuracy of   $10^{-4}$~s for the next generation of detectors \footnote{This accuracy is conservative compared with an estimate based on the uncertainty on the vertex reconstruction, which is about 3 m for Hyper-Kamiokande \cite{Abe:2011ts}. In order to obtain a global time, when comparing with other experiments, a higher uncertainty is expected.},  the time lapse differences between NO and IO will not be distinguishable.
  \begin{figure*}[t]
  	\subfigure
  	{\includegraphics[width=8cm]{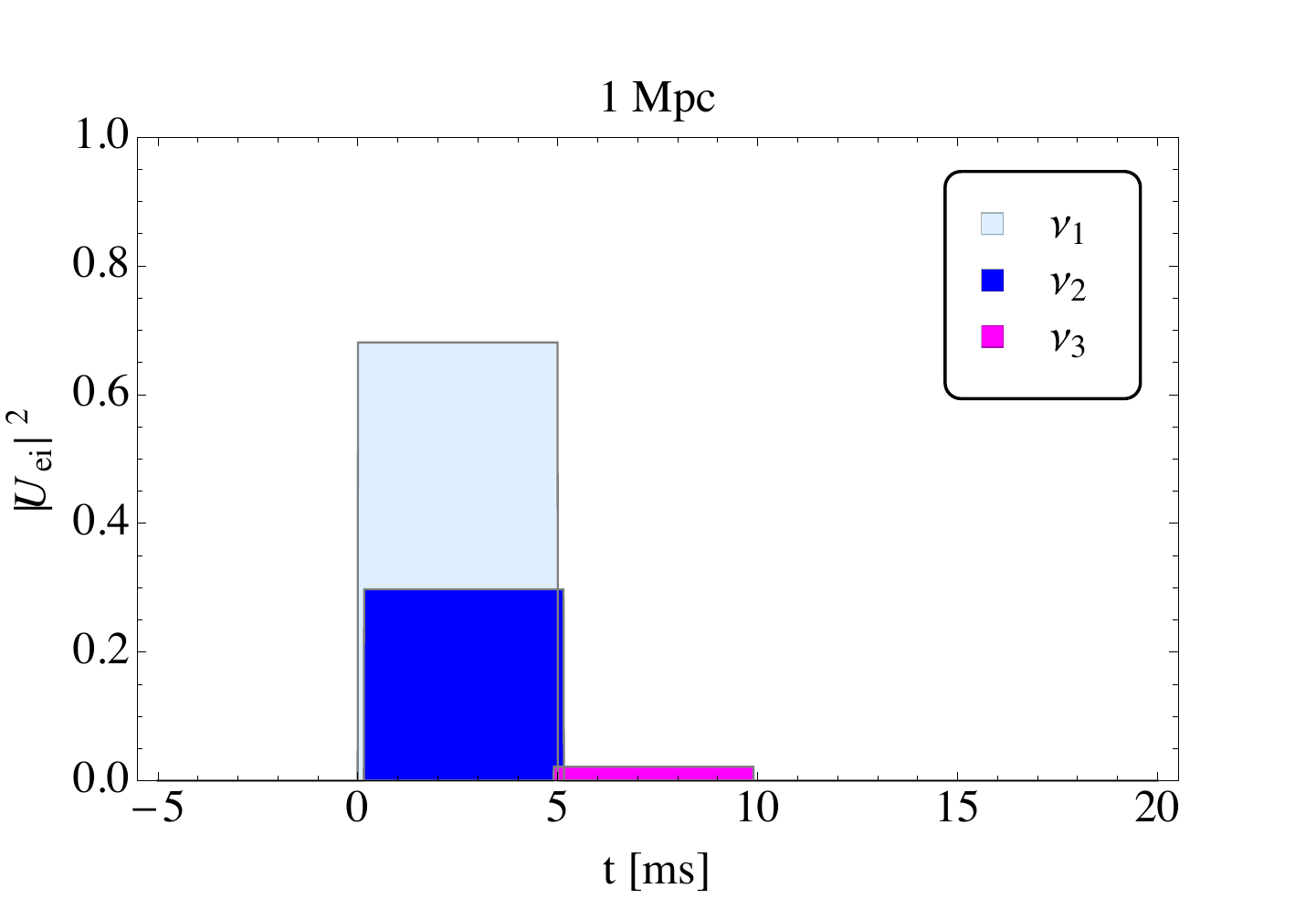}}
  	\subfigure
  	{\includegraphics[width=8cm]{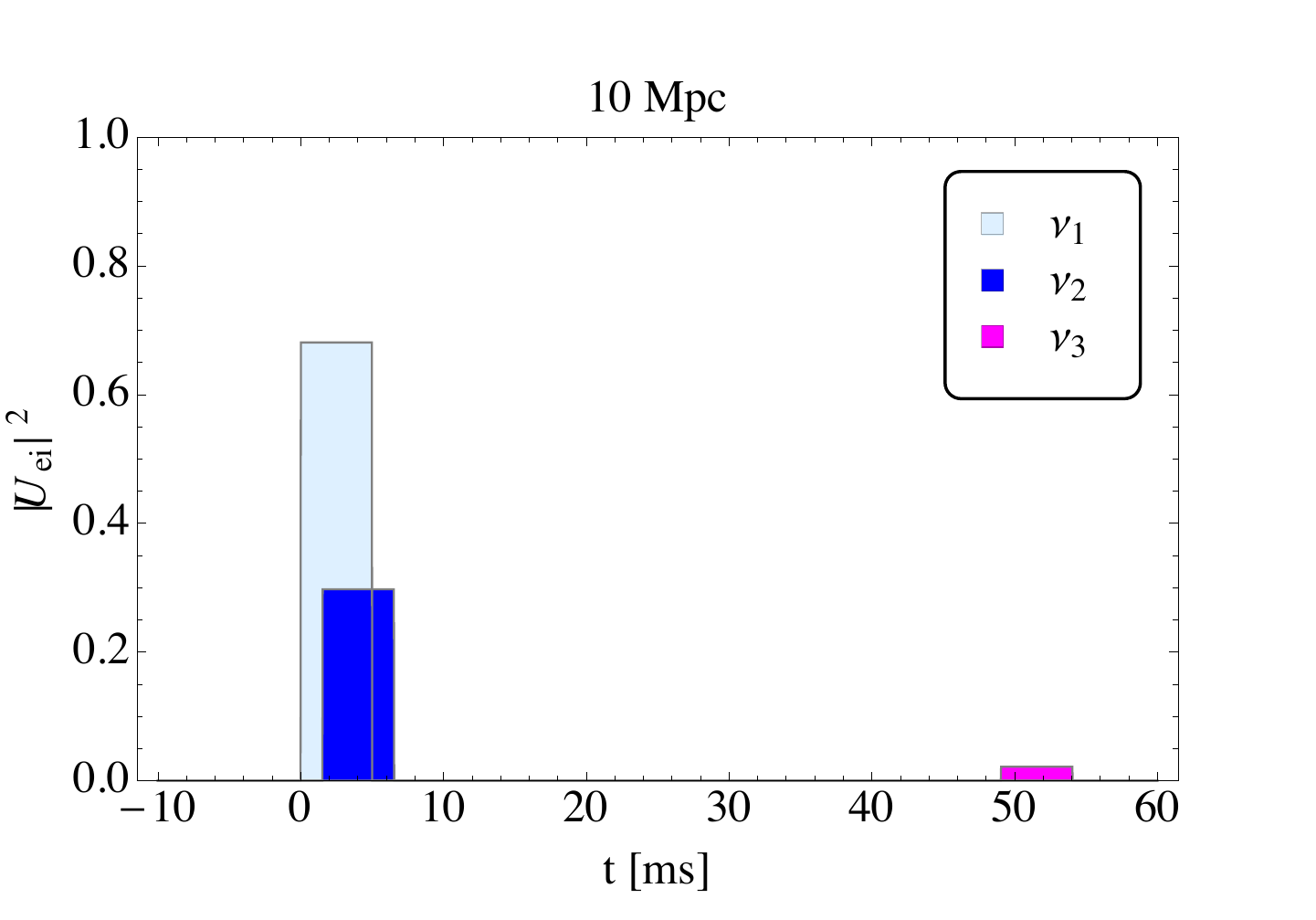}}
  	\vspace*{-10pt}\caption{\label{fig:probplot} Schematic representation of the square root of the probability given in (\ref{ICprobi}) of detecting flavor state $\nu_e$ if the source emits a short burst of $\nu_e$ as a function of time. The left panel is for 1 Mpc and the right, for illustrative purpose,  is for 10 Mpc at an energy of 5~MeV. For definiteness we assume NO and each bin corresponds to a fiducial collective time of 5~ms. }
  \end{figure*}
  
However, in addition to the time information, also the ratio between the amplitudes
of the different neutrinos reaching the detector can be measured. Since the distances considered here are very large, neutrinos will reach the detector incoherently such that the time integrated arrival probability is:
\be P\left(\nu_\alpha\rightarrow \nu_\beta\right) =\sum_i \left| U_{\alpha i }\vphantom{U_{\beta i}}\right|^2 \left| U_{\beta i }\right|^2,\ee
where $\alpha$ and $\beta$ are flavour eigenstates.
In fact, this expression  holds true whenever the  time arrival differences among the three mass eigenstates  is smaller than the detector time resolution. However, when $\Delta t_{\nu_i\nu_j}$ is larger than the detector resolution, then each mass eigenstates $\nu_i$ can be detected independently and will interact with the detector with probability \footnote{We work in the regime of incoherence. Defining  $\sigma_{xP}$ ($\sigma_{xD}$) as the spatial width of the production (detection) neutrino wave packet, we work under the assumption that 
$|(v_j -v_k) L/c | \gg {\rm max}(\sigma_{xP},\sigma_{xD})$ being $v_i$ and $v_j$ the  two group velocities of the two wave packets of neutrino mass eigenstates $\nu_i$ and $\nu_j$.}
 \be\label{ICprobi}
 P\left(\nu_\alpha\rightarrow \nu_\beta\right)_i =  \left| U_{\alpha i }\vphantom{U_{\beta i}}\right|^2\left| U_{\beta i }\right|^2 \ . \ee
For simplicity, here we do not consider matter effects which could in principle take place in the propagation through the Earth itself.
In Fig.~\ref{fig:probplot} we illustrate a possible pattern of neutrino detection following \eqref{ICprobi} using also the time-differences reported in Table~\ref{tab:BP5MeV}. Depending on the source, its distance and the experimental time sensitivity the figure shows that, at least in principle, one can observe interesting time-patterns reflecting the neutrino ordering and mixing.\\
So far we discussed the basic setup and argued that neutrino detectors on Earth can help disentangle the neutrino ordering, when observing distant astrophysical catastrophes.   It is time to move to additional precious information that we can gain when comparing time differences with respect to the other light messengers.
  
\subsection{Absolute neutrino masses from time differences}
The attractive idea to use a multi-signal approach  was put forward in  
\cite{Nishizawa:2014zna} where the authors translate a potential SN signal of GWs and neutrinos into 
 limits on the  speed of GWs and on the absolute neutrino mass scale. 
 We define: 
\be 
\Delta T_{\nu_i g} = T_{\nu_i}-T_g \ , 
\label{DT}
\ee
 that implies:
\be \Delta T_{\nu_i g } = \tau_{obs}^i -  \tau_{int}^{\nu}\label{eq:deltag},\ee  
where $i$ denotes now the $i$-th  neutrino mass eigenstate.
The deviation from the speed of light for GWs and neutrinos reads: 
    \be \delta_g\equiv \frac{c-v_g}{c}, \qquad \delta_{\nu_i}\equiv \frac{c-v_{\nu_i}}{c},\ee
with:      \be \delta_{\nu_i}= \frac{m_{i}^2 c^4}{2 E^2}+  \mathcal{O}\left(\frac{m_i^4 c^8}{8E^4} \right).\ee
  From the definition \eqref{DT} follows:   
    \be \frac{ \Delta T_{\nu_i g}}{T_0}= \frac{\delta_{\nu_i}-\delta_g}{(1-\delta_g)(1-\delta_{\nu_i})},\label{eq:del}\ee
  where, as already defined earlier, $T_0=L/c$. 
 If  in  \eqref{eq:del} we consider an uncertainty in the time of emission of neutrinos, $\tau_{int}^\nu$,  in order to detect the GW and the neutrino signal, we must have:
\be|\Delta T_{\nu_i g}| > \tau_{int}^\nu,\ee
and using  \eqref{eq:del} to the first oder in $\delta_\nu$ and $\delta_g$ one finds:
\be  | \delta_{\nu_i}-\delta_g |T_0  \gtrsim \tau_{int}^\nu\label{eq:limit}.\ee
 \begin{figure}[t]
 	{\includegraphics[width=7.5cm]{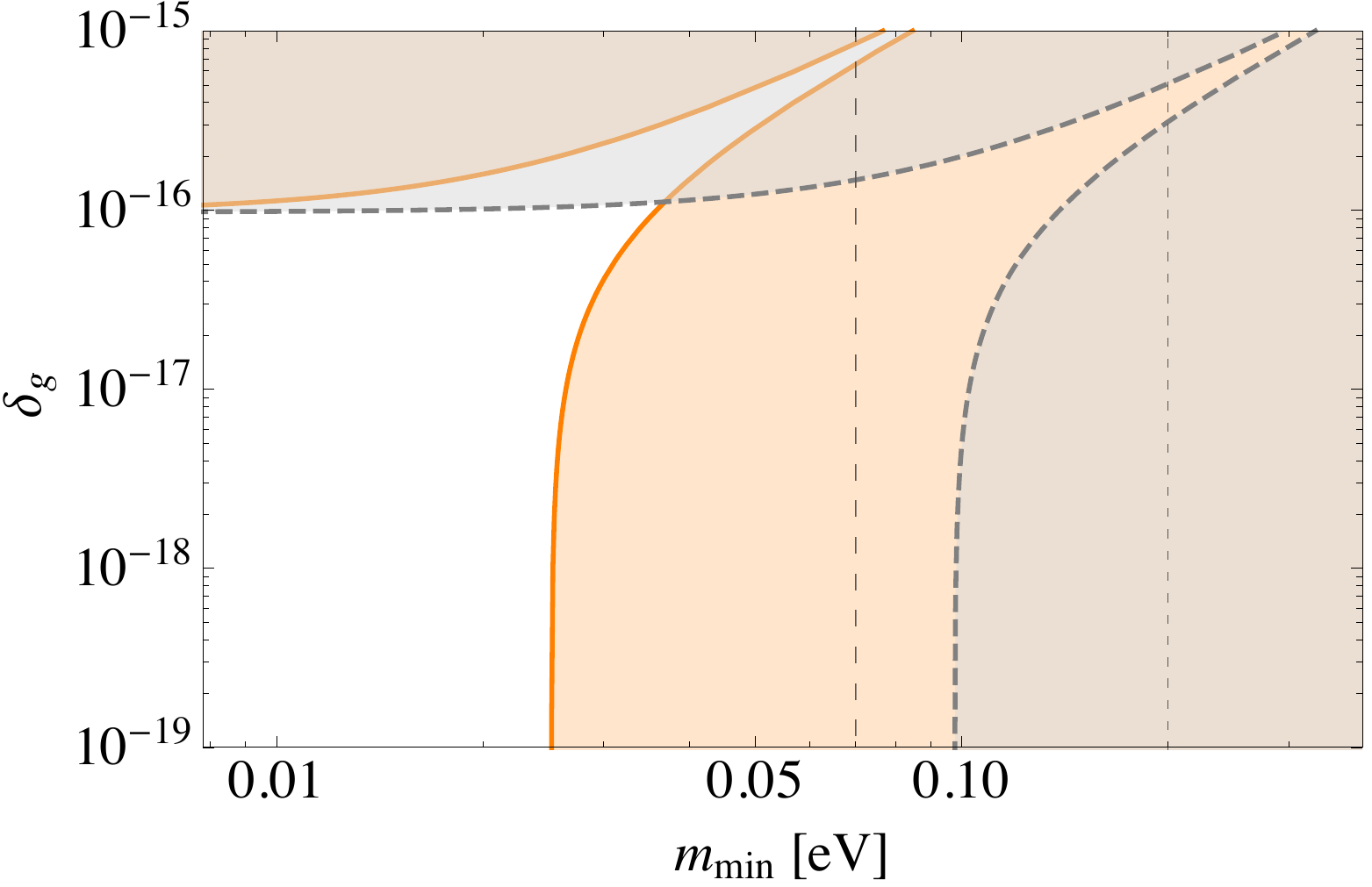}}
 	\vspace*{-10pt}\caption{\label{fig:deltag1}  Plot of $\delta_g$ as function of the lightest neutrino mass considering $\tau_{int}^\nu\sim 10$~ms considering the  eq. \eqref{eq:limit}. We consider of the HK and JUNO energy thresholds, $E_\nu=7$~MeV (grey region) and $E_\nu=1.806$~MeV (orange region), respectively, for  $L=  1$ Mpc. 
 		The dashed and  dotted  vertical lines correspond  to the Planck limit,  {$ \sum_i m_i <0.23$}~eV,  and the perspective upper limits of KATRIN,   ~{$0.2$}~eV. }
 \end{figure}
 Using the inequality above and assuming $\tau_{int}^\nu\sim 10$~ms (typical time for a SN burst) and an energy equal to the energy threshold of HK, $E_\nu=7$~MeV, we show in Fig.~\ref{fig:deltag1}  the $\delta_g$ dependence on the lightest neutrino mass for a reference distance of $L=  1$ Mpc (grey region). 
In principle, detectors with a lower energy resolutions, such as JUNO ($E^{th}_\nu$=1.806 MeV) could test   lower values of $m_{min}$ and could probe neutrino mass up to $\sim 0.02$ eV for distances around 1 Mpc, which are at least an order of magnitude lower than present  cosmological limits and the perspective upper limit from KATRIN, see orange region in Fig.~ \ref{fig:deltag1}.  
 \\
 Last, we notice that limits on $v_g$ can also be obtained from high energetic events or from the requirement of Lorentz invariance.
In fact, if the GW velocity is subluminal, then cosmic rays lose their energy via gravitational Cherenkov radiation and cannot reach the Earth. The fact that ultra-high-energy cosmic rays are observed on  Earth limits the GW propagation speed to be 
\be c - v_g < 2 \times 10^{-15} (10^{-19}) c, \label{eq:CR}\ee 
assuming that the cosmic rays have galactic origin (extra-galactic) \cite{Moore:2001bv}.  
\\
Further independent constraints on Lorentz violation can therefore be set when observing photon and gravitational waves. An attempt of doing so  appeared in \cite{Ellis:2016rrr}  
 by combining the event GW150914 in GWs with the observation made by the Fermi Gamma-Ray Burst Monitor \cite{Fermi-LAT:2016qqr} of a transient photon source in apparent coincidence  \cite{Ellis:2016rrr}  
 \be v_g-c< 10^{-17} c.\label{eq:LIGO}\ee
There are serious concerns about the true correlation between the two events. Nevertheless if one recasts the limits in eqs. \eqref{eq:CR}  and \eqref{eq:LIGO} one obtains:
  \be - 10^{-17}<\delta_g < 2 \times 10^{-15} (10^{-19}).\ee  
Independent bounds on $\delta_g$ are important, since they allow for a more precise interpretation of Fig.~\ref{fig:deltag1} in terms of $m_\text{min}$. 

We discussed so far the time difference measurement between a neutrino and a GW. Similarly one can imagine a time difference to emerge if rather than a GW, one were to detect a photon. If all messengers were simultaneously detected and assuming a unique source  by using  \eqref{eq:TOF}, within experimental resolution, the following consistency condition must hold:
{\be \Delta t_{\gamma \nu_i }-\Delta t_{\gamma \nu_j}=  \Delta t_{\nu_j \nu_i}=   \Delta t_{g \nu_i }-\Delta t_{g \nu_j}\ .\ee

\section{Concluding with a Preliminary Feasibility Study}
So far we have been concerned with the theoretical setup, and since the framework presented here relies on distant sources, we will now perform a preliminary  study of the actual experimental feasibility. In the following, we will not discuss the distribution and the expected number of various kinds of astrophysical events,  but focus on the number of detected neutrinos assuming a specific source at a given distance.
From the analysis above it is clear, that three parameters are vital to increase the time lapse between mass eigenstates: the distance from the source $L$,  the energy of the emitted neutrino, $E_\nu$, and the 
absolute neutrino mass $m_{min}$. Conversely, the larger the distance is, the smaller is the rate.  As a consequence, if the neutrino counterparts of events  like GW150914 would be emitted by the source, it would be hard, if not impossible, to detect them on Earth. \\
As a benchmark investigation we will concentrate on the next generation of neutrino detection experiments such as  1 Mton   Hyper-Kamiokande (HK)  in Japan  \cite{Abe:2011ts} that has already sparkled interests in the astrophysical community. 
Astrophysical catastrophes  like the merging of a  neutron star black hole binary or the core bounce of a core-collapsed supernova are expected to produce a total neutrino output carrying an overall energy of circa $10^{53}$ erg. For such an event one expects on Earth an integrated time flux per squared meter of about $3\times10^{11}\left( d /\text{Mpc }\right)^{-2}\,m^{-2}$. Despite the fact that a large number of neutrinos will reach Earth  because of their low cross section only a tiny fraction will be detected. Previous studies \cite{Ando:2005ka} indicate that HK can detect 1-2 neutrino events per year from supernovae in the range up to 10 Mpc. 
\\
However, our theoretical analysis made use only of the neutrinos emitted during the initial burst from the source which can be determined by integrating the following neutrino detection rate over the relevant time interval: 
\be \frac{dN}{dt}=n_p\int_{E^{th}_e} dE_e  \int_{E^{th}_\nu} \,dE_\nu \,\mathcal{F}(E_\nu, t)\,\sigma^\prime(E_e,E_\nu)\,\epsilon,\ee
where $n_p$ is the number of protons in the target,  $E_{\nu, e}$ are respectively the (anti)neutrino and the (electron) positron energy of the event, $\mathcal{F}(E_\nu, t)$ is the flux per unit time, area and energy and $\epsilon$ is the detector efficiency. Finally $\sigma^\prime(E_e,E_\nu)=d\sigma/dE_e$ is the differential cross section of the process under study.
 We will assume the efficiency of the detector to be 100\% for energies larger than the energy threshold  of the detector, $E_\nu> E^{th}_\nu$. 
\\
Our estimates assume a typical energy in neutrinos emitted from astrophysical sources within the initial burst to be of the order of $\sim 10^{51}$ erg as well as a mean neutrino energy $\langle E_{\bar \nu_e}\rangle \sim 12 $ MeV.   From a SN at a distance $d$, HK (0.74 Mton, $E^{th}_{\nu}$=7~ MeV, $E^{th}_{e}$=4.5~MeV) would expect the following number of detected neutrinos (indicated by $\lambda_{\text{ES}}$) via  neutrino-electron elastic scattering (ES)  processes:
\be\label{SN}
\lambda_{\text{ES}} = 1.8\times 10^{-3} \, \left( \frac{d}{\text{Mpc}}\right)^{-2}
\ee
 where the initial burst is primarily $\nu_e$ from the neutronization process. Similarly,  from  a neutron star black hole (NS-BH) merger, where the burst consists mostly of $\bar{\nu}_e$, we get via inverse beta decay (IBD) \cite{Strumia:2003zx} a number of neutrinos of
\be\label{merger}
\lambda_{\text{IBD}} = 1.6\times 10^{-1}  \, \left( \frac{d}{\text{Mpc}}\right)^{-2}.
\ee\\
For such low rates it is useful to estimate the actual detection probability as function of the distance from the source. To assess this, we use the Poisson probability to detect $n$ events  as  
$P_n= \lambda^n e^{-\lambda}/n!$ where $\lambda$ is the expected number of events, given in eq. \eqref{SN} or eq.  \eqref{merger}. In Fig.~\ref{fig:Pdet} we show, as an illustrative example, the detection probability for IBD resulting from requiring at least one, two, and 10 events per burst, indicated respectively with blue, red and black curves. We use in our estimates the energy range $7-30$~MeV. The plot shows  the HK detection probability for $\bar\nu_e$ for a NS-BH merger (solid line), as well as the one for a hypothetical 5 Mton detector (dashed line), e.g. \cite{Suzuki:2001rb}. We observe that even for $\sim$1~Mpc and a $7-30$ MeV energy range one can still observe $\sim1$ event. These  estimates show that it is possible to reach phenomenologically interesting neutrino mass differences from sources at $\sim$1 Mpc provided one can combine more than one Mton experiment. 
In order to compute the expected annual rate of detected neutrino events, one has to combine the above analysis with the annual rate of relevant astrophysical events. The annual rate of SNs is expected to be $1/3\, \mathrm{yr}^{-1}$ within 4 Mpc \cite{Ando:2005ka}, while the rate for NS-BH mergers is more uncertain with an expected rate of $10^2-10^3 \mathrm{yr}^{-1}$ within 1 Gpc. This rate will in the future be constrained by LIGO \cite{Abbott:2016ymx}.\\
 We stress that we have used conservative estimates, for example, in the total energy emitted with the neutrino burst. Another parameter that can be played with is the time resolution in neutrino detection that can, in the future, be expected to go below one millisecond. If this is the case it would allow sources as close as 100 kpc to become relevant for our analysis.  In this case the neutrino flux increases by two orders of magnitude.
 \begin{figure}[t]
  {\includegraphics[width=7.5cm]{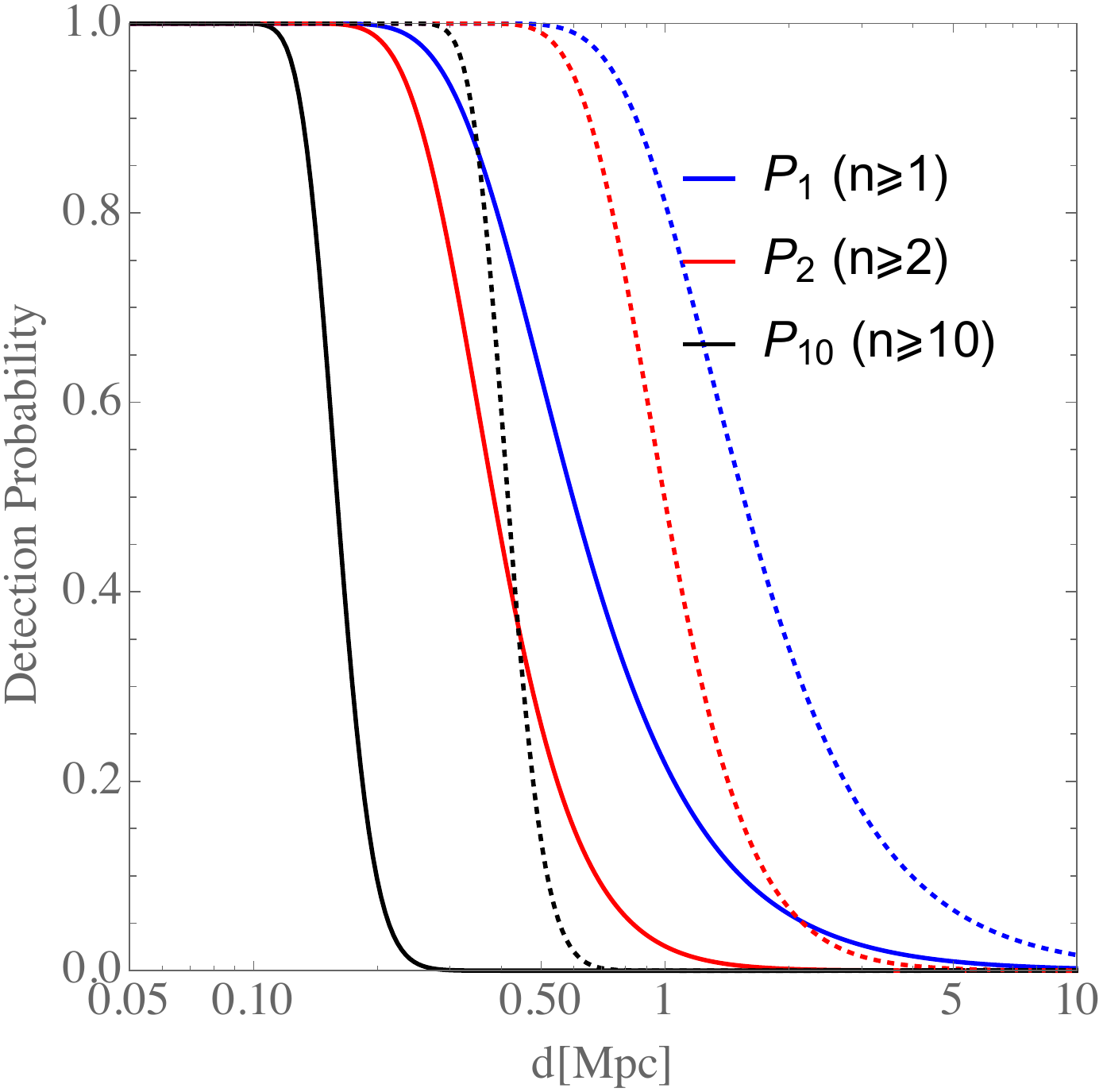}}
      \vspace*{-10pt}\caption{\label{fig:Pdet} Detection probability of  neutrinos versus distance from the source to Hyper-Kamiokande (solid lines) and to a hypothetical future 5 Mton experiment (dotted lines) \cite{Suzuki:2001rb} using a $7-30$ MeV energy range. Blue, red and black curves represent the detection probability resulting in requiring observation of at least one, two, and ten events per burst, respectively.    }
 \end{figure}
\\
\\
\indent  To conclude, we derived the theoretical and phenomenological conditions under which multi-messenger astronomy can disentangle or further constrain the neutrino mass ordering. We have also argued that it can provide salient information on the absolute neutrino masses.  We added a  preliminary feasibility study to substantiate and further motivate our theoretical analysis. We have seen that future experiments can be useful also in testing independently the cosmological bounds on neutrino absolute masses. However, this requires high resolution timing and a significant increase in the combined fiducial volume compared to the current Cherenkov water detectors. 
 
  Conversely one can use future results on neutrino properties to provide detailed information about astrophysical sources emitting simultaneously GWs, photons and neutrinos, and possibly lower uncertainties in the emitted multi-messenger signal from the source. \\    
  
  \section*{Note added in proof} While our work was under review related papers on the propagation time of ultra-relativistic particles appeared in the literature \cite{Fanizza:2015gdn,Fleury:2016mul}, which provide relevant details for a high precision application of the presented framework.

\section*{Acknowledgments}
 The CP3-Origins centre is partially funded by the Danish National Research Foundation, grant number DNRF90.  
%
\bibliography{}
%

\end{document}